\documentclass{aa}  
\usepackage{graphicx}
\usepackage{natbib}
\usepackage{txfonts}
\usepackage{multirow}
\begin{document}
   \title{X-ray activity cycle on the active ultra-fast rotator AB~Dor~A?
\subtitle{Implication of correlated coronal and photometric variability}}

   \author{ S. Lalitha and J. H. M. M. Schmitt
          }
   \authorrunning{S. Lalitha et al.}
   
   \offprints{S. Lalitha}

   \institute{Hamburger Sternwarte, University of Hamburg,
              Gojenbergsweg 112, 21029 Hamburg, Germany\\
              \email{lalitha.sairam@hs.uni-hamburg.de}
               }

   \date{Received XXXX; accepted XXXX}

 
\abstract
   {Although chromospheric activity cycles have been studied in a larger number of 
late-type stars for quite some time, 
very little is known about coronal activity-cycles in other stars and 
their similarities or dissimilarities with the solar activity cycle.}
{ While it is usually assumed that cyclic activity is present only in stars of low to moderate activity,
we investigate whether the ultra-fast rotator AB~Dor, a K dwarf exhibiting 
signs of substantial magnetic activity in essentially all wavelength bands,
exhibits an X-ray activity cycle in analogy to its photospheric activity cycle of about 17 years 
and possible correlations between these bands.}
   { We analysed the combined optical photometric data of AB~Dor~A, which span $\sim$35 years. Additionally, 
   we used ROSAT and \emph{XMM-Newton} 
X-ray observations of AB~Dor~A to study
the long-term evolution of magnetic activity in this active K dwarf over nearly three decades
and searched for X-ray activity cycles and related photometric brightness changes. }
{ AB~Dor~A exhibits photometric brightness variations ranging between $6.75~<~V_{mag}~\leq~7.15$ while 
the X-ray luminosities range between $29.8~<~log~L_X~[erg/s]~\leq~30.2$ in the 0.3-2.5~keV.  
As a very active star, AB~Dor~A shows frequent X-ray flaring, but, in the long \emph{XMM-Newton} 
observations a kind of basal state is attained very often. This basal state 
probably varies with the photospheric activity-cycle of AB~Dor~A which has a period of $\sim$17~years, but, the X-ray 
variability amounts at most to a factor of $\sim$2, which is, much lower than the typical
cycle amplitudes found on the Sun.} 
{}

   \maketitle
%

\section{Introduction}

One of the key characteristics of the Sun is its 11-year activity cycle, 
which was originally discovered from the periodic variation in the observed 
sunspot numbers \citep{schwabe_1844}. 
However, the solar cycle also manifests itself in 
many other activity indicators such as the solar 10.7 cm radio emission, its 
chromospheric \ion{Ca}{ii} emission, and its coronal X-ray emission \citep{gnevyshev_1967,hathway_2010}.  
Ever since \cite{hale_1908} discovered strong magnetic fields in sunspots, the
magnetic character of solar activity and its cyclic variations has been beyond dispute.
The cycle variations in the solar corona are far more pronounced than those observed
in the photosphere, and  
with the advent of space-based astronomy vast amounts of solar X-ray data 
have been collected, which allow a better understanding of the evolution of coronal plasma 
temperature, emission measure, and structure over the solar cycle \citep{orlando_2000}. 
As expected, the solar corona has a distinctly different appearance during activity minimum 
than encountered at activity maximum. The solar X-ray flux varies during a cycle by typically a 
factor of $\sim 200$ in the energy range of 0.6-1.5~keV \citep{kreplin_1970}, while
\cite{stern_2003} computed the solar soft X-ray irradiance variations as 
measured by the \emph{Yohkoh} satellite in the 0.5-4~keV energy range
and found a maximum- to minimum- ratio of $\sim 30$. \cite{tobiska_1994} estimated 
a variation of a factor of $\sim$10 in the softer energy range between 0.25-0.4~keV as the solar activity cycle progresses. 
The cause for these substantial flux variations is
the absence of large active regions in the solar corona during its activity 
minimum, which make it appear much fainter than the solar corona during the activity maximum \citep{golub_1980}. 
However, the amplitude of these variations sensitively depends 
on the energy range considered and becomes much smaller at softer energies.  
For example, \cite{ayres_1997} and \cite{ayres_2008} argued that through the 
0.1-2.4~keV \emph{ROSAT} pass band, the Sun was expected to show variations between the minimum and maximum flux of a factor of 5-10.
\cite{peres_2000} estimated the X-ray brightness of the Sun during   
activity maximum and minimum to be log~L$_X$~=~27.5~erg/s and log~L$_X$~=~26.5~erg/s, respectively, in the 
ROSAT 0.1-2.4~keV energy band, which would mean a variation of order of magnitude variation in the ROSAT band. 
Their results were supported by the studies of \cite{orlando_2000} 
and \cite{judge_2003}, who also estimated variations of about a factor of $\sim$10 in the solar 
coronal X-ray emission throughout the solar cycle.

The question then immediately arises whether other late-type stars also show such a 
solar-like cyclic variability in their magnetic activity properties \citep{vaughan_1978, wilson_1978}. 
Within the context of the Mt.~Wilson HK program the long-term variability of stellar 
chromospheric activity (as observed in the Ca~II emission cores) in a large sample of 
late-type stars was systematically studied over several decades and the cyclic activity of a 
larger number of stars in the solar neighbourhood was established \citep{baliunas_1995}. 
Specifically, \cite{baliunas_1998} found that about 60\% of the stars of the 
Mt. Wilson observatory survey exhibited periodic and cyclic variations, and furthermore, 
\cite{lockwood_2004, lockwood_2007} found evidence that the photospheric and
chromospheric activity cycle are related. 
\cite{baliunas_1995} employed the same technique of studying stellar activity 
to also monitor the solar activity-cycle in integrated light and showed that 
the so-called S-index of the Sun varies between 
0.16 and 0.22 between its activity minima and maxima; in fact, the Sun shows one of the most regular cycles
in the whole stellar sample presented by \cite{baliunas_1998}.

A related question is whether the stars with different cyclic properties in their Ca~II emission
show a different behaviour with respect to their coronal emission.
\cite{hempelmann_1996} compared soft X-ray 
fluxes with Ca emission in a sample of late-type stars
and showed that the stars with cyclic variations in their calcium flux tend to show less 
X-ray activity than stars with irregular variability in their Ca~II emission. 
Additionally, X-ray faint stars tend to show flat activity curves 
or low levels of short-term variability (see \citealt{wright_2010} and references therein). 
On the other hand, X-ray bright, active stars are believed to have no long-term cycles; 
instead, they are thought to exhibit an irregular variation in their X-ray luminosity \citep{stern_1998}. 
Currently, only fewer than handful of stars have been found to have long-term X-ray cycles (cf. Fig.\ref{cycrot}). 
With ROSAT, the monitoring of the visual binary 61 Cyg began that subsequently was continued with \emph{XMM-Newton} \citep{hempelmann_2006}. 
Four more stars, which are the $\alpha$ Cen system, HD 81809, and $\tau$ Boo, were also monitored by 
XMM-Newton \citep{ favata_2004,favata_2008, robrade_2007, ayres_2009, robrade_2012, poppenhaeger_2012} for possible cyclic variations.
\cite{favata_2004} detected a pronounced cycle of 8.2 years and a 
clear evidence for large-amplitude X-ray variability in phase with the chromospheric activity 
cycle for HD 81089. 
For 61 Cyg A, \cite{robrade_2012} found a regular coronal 
activity cycle in phase with its 7.3 yr chromospheric cycle, whereas no evidence of a 
clear coronal cycle for 61 Cyg B could be produced. Furthermore, these authors demonstrated that the 
two $\alpha$ Cen stars exhibit significant long-term X-ray variability, with 
$\alpha$ Cen A showing a cyclic variability over a period of 12-15 years, while
the $\alpha$ Cen B data suggest an X-ray cycle of a period of 8-9 years;
the amplitudes of the variability for $\alpha$ Cen A and B were 
estimated to be an order of magnitude and about a factor six to eight, respectively. 
In addition, Robrade and collaborators also concluded that the coronal activity cycles are a common phenomenon in older, 
slowly rotating G and K stars. It is worthwhile noting that most of these stars were moderately or low active. However, 
recently \cite{sanz_2013} showed an X-ray cycle of $\sim$1.6~years 
in the active planet-hosting star $\iota$ Hor, demonstrating that short cycles in Ca~II also have an X-ray equivalent.
On the other hand, \cite{poppenhaeger_2012} studied 
the activity cycle associated with $\tau$ Boo, 
a moderately active F-star displaying a magnetic cycle of $\sim$ 1 year, as 
anticipated from Zeeman Doppler imaging, 
but they were unable to find any evidence of an activity cycle with the 
available X-ray data.

We study the behaviour of coronal X-ray emission during the activity cycle and 
a possible correlation between the photospheric and X-ray 
activity in the very active star AB~Dor~A.
Our paper is structured as follows: in Sect.~\ref{sec:target} 
we present the target stars and in Sect.~\ref{sec:obsdata} we describe the observations and data analysis. 
In Sect.~\ref{sec:lc} we discuss the optical and X-ray light curves and also discuss our 
investigation on the correlation between the photospheric and coronal activity cycles.  In Sect.~\ref{sec:compare}, we compare 
our target with other stars that show cyclic behaviour. 
In Sect.~\ref{sec:rotmod} we investigate the short-term variation induced by the star's rotation, and 
we close with a summary in Sect.~\ref{sec:sum}.

\section{Our target star}\label{sec:target}

AB~Dor is a quadruple system consisting of the components 
AB~Dor~A, AB~Dor~Ba, AB~Dor~Bb, and AB~Dor~C.
AB~Dor~A is a magnetically active young dwarf-star of spectral type K0, located 
at a distance of $\sim$15 pc from the Sun as a foreground star of the
Large Magellanic Cloud (LMC). It is a very rapid rotator with a 
period of $P$ = 0.514 days and $v~sin$i~$\approx$~90~Km/s 
(see \citealt{guirado_2011} and references therein), 
resulting in very high levels of magnetic activity with an average 
$log(L_{x}/L_{bol}) \approx $-3. Located 9.5$\arcsec$ away from 
AB~Dor~A is an active M dwarf AB~Dor~B  (Rst~137B; \citealt{vilhu_1987,vilhu_1989}),
about $\sim$60 times bolometrically fainter than AB~Dor~A, and therefore 
only little or no contamination due to the presence of AB~Dor~B is expected in data that
leave both components unresolved. 
At radio wavelengths AB~Dor~B was serendipitously detected with the Australian 
Telescope Compact Array (ATCA) during an observations of AB~Dor~A \citep{lim_1992}. 
However, the binarity of AB~Dor~B itself with a separation of only 0.7$\arcsec$ 
(called AB~Dor~Ba and AB~Dor~Bb) was detected only after the advent of adaptive optics. 
Yet another low-mass companion to AB~Dor~A is AB~Dor~C \cite{guirado_1997}, located about 
0.16$\arcsec$ away from AB~Dor~A.

The apparent magnitude of AB~Dor~A of V=6.75 
\citep{amado_2001} makes it a favourite target for optical observations with the aim of 
monitor photospheric spots and performing Doppler imaging \citep{rucinski_1983, innis_1986, innis_1988, 
kuerster_1994, anders_1994, unruh_1995}. 
Additionally, \cite{jarvinen_2005} noted evidence for a possible 
activity cycle of $\sim$20 years along with a flip-flop cycle of $\sim$5.5 years. 
\cite{innis_2008} repeated the cycle study with new data and determined a cycle period 
supporting the $\sim$20 year period suggested by \cite{jarvinen_2005}.

AB~Dor~A has not only been a target of interest for optical observations, 
but has been observed with many space-based observatories across the UV, EUV, and X-ray wavebands. 
The first X-ray detection of AB~Dor~A  was obtained with the \emph{Einstein} Observatory 
\citep{pakull_1981, vilhu_1987}, and 
ever since then AB~Dor~A has been observed repeatedly by almost all X-ray observatories 
\citep{collier_cameron_1988, vilhu_1993, mewe_1996, maggio_2000, guedel_2001, 
sanz_forcada_2003, hussain_2007, lalitha_2013}. The long-term X-ray behaviour of
the X-ray emission from the AB~Dor system is dominated by AB~Dor~A \citep{guedel_2001, sanz_forcada_2003}. 
AB~Dor~Ba and Bb cannot be separated with current X-ray telescopes;
the combined luminosity of the B-components is $\sim2.8\times10^{28}$~erg/s in the 0.2-4.0~keV \citep{vilhu_1987}. 
\cite{sanz_forcada_2003} obtained a luminosity of 
$\sim3.4\times10^{28}$~erg/s in the 0.5-2.0~keV with \emph{Chandra} ACIS observations. 
Hence, the contribution of the companions 
to the X-ray emission of AB~Dor~A can be considered negligible, essentially because the 
quiescent X-ray emission of the companions scales as their bolometric luminosity. 

The time evolution of AB~Dor~A has previously been studied by \cite{kuerster_1997}, who compared 
the V-band brightness with X-ray observations (5~1/2 years of observations) carried out by the ROSAT satellite, 
but they found no pronounced long-term activity period from their analysis because the 
5~1/2 years of data available at the time barely cover a part of the activity cycle.  The same applies
to the studies of \cite{sanz_2007}, who noted a weak increasing trend in 
the X-ray emission using the observations carried out by \emph{XMM-Newton} that cover about six  years of observations.

\section{Observations and data analysis}\label{sec:obsdata}

\subsection{X-ray data}

Because it is a foreground star of the LMC, AB~Dor has the advantage
of being easily observable at all times with most X-ray satellites, and therefore quite a number
of often serendipitously taken X-ray data of this source exist. 
We used 
\emph{ROSAT} observations\footnote{The ROSAT observation log is provided in electronic form at CDS.} 
listed in the \emph{ROSAT} Position Sensitive Proportional Counters (PSPC) source catalogue
from pointed observations with typical exposure times of between 1~ksec and 3~ksec,
and the \emph{ROSAT} High Resolution Imager (HRI) source catalogue again from pointed observations with 
typical exposure times of between 1~ksec and 6~ksec. Since the ROSAT satellite 
was in a low Earth orbit, the typical contiguous and uninterrupted viewing 
intervals of a source are typically in the range 1~-~2 ksec, therefore longer 
exposures are composed of a number of shorter exposures with sometimes very long 
intervening temporal gaps. 
We specifically used the PSPC observations obtained between 1990 and 1993
and the HRI observations obtained between 1990 and 1998; the total PSPC exposure is 
74.4 ksec, the total HRI exposure is 106.2 ksec; thus, the ROSAT observations comprise a 
relatively short total exposure time when compared with the \emph{XMM-Newton} observations 
listed in Table~\ref{obs}. 

We also carried out a detailed analysis of AB~Dor~A, using the data obtained
by \emph{XMM-Newton} Observatory. 
On board {\em XMM-Newton} three telescope are co-aligned with three CCD 
cameras (i.e., one PN and two MOS cameras) with a sensitivity range between 
$\approx$~0.2~and~15~keV, which together form the European Photon Imaging Camera (EPIC). 
The X-ray telescopes equipped with MOS detectors are also equipped with reflection 
gratings. These two reflection-grating spectrometers (RGS) provide high spectral resolution 
(E/$\Delta$E $\approx$ 200-800) in the energy range 0.35-2.5~keV. Useful data were 
obtained from the EPIC and the RGS detectors (see Tab.~\ref{obs} for a detailed account). 

AB~Dor~A, which is a very bright target with many emission lines, is fortuitously used
as a calibration source for the \emph{XMM-Newton} RGS. Hence this 
target has been repeatedly observed over the last decade, giving us an ideal 
opportunity to assess the long-term behaviour of AB~Dor~A. In these data
there are either no observations or typically much shorter observation time covered by the EPIC 
instrument than that of the RGS (see Tab.~\ref{obs}); 
we therefore restricted our analysis to the available RGS data. 
The data were reduced using the standard {\em XMM-Newton} Science Analysis System (SAS) software V12.0.1. 
We used the meta-task \emph{rgsproc} 1.30.3 to process the RGS data, 
followed by the spectral extraction and response generation. 
To create a combined light curve of the two instruments 
(RGS1+RGS2) the task \emph{rgslccor} 0.52.1 was used\footnote{
A detailed description of the 
XMM packages is available at http://xmm.esac.esa.int/sas/current/doc/packages.All.html}. 

\begin{figure*}[!ht]
\begin{center}
\includegraphics[width=16.5cm]{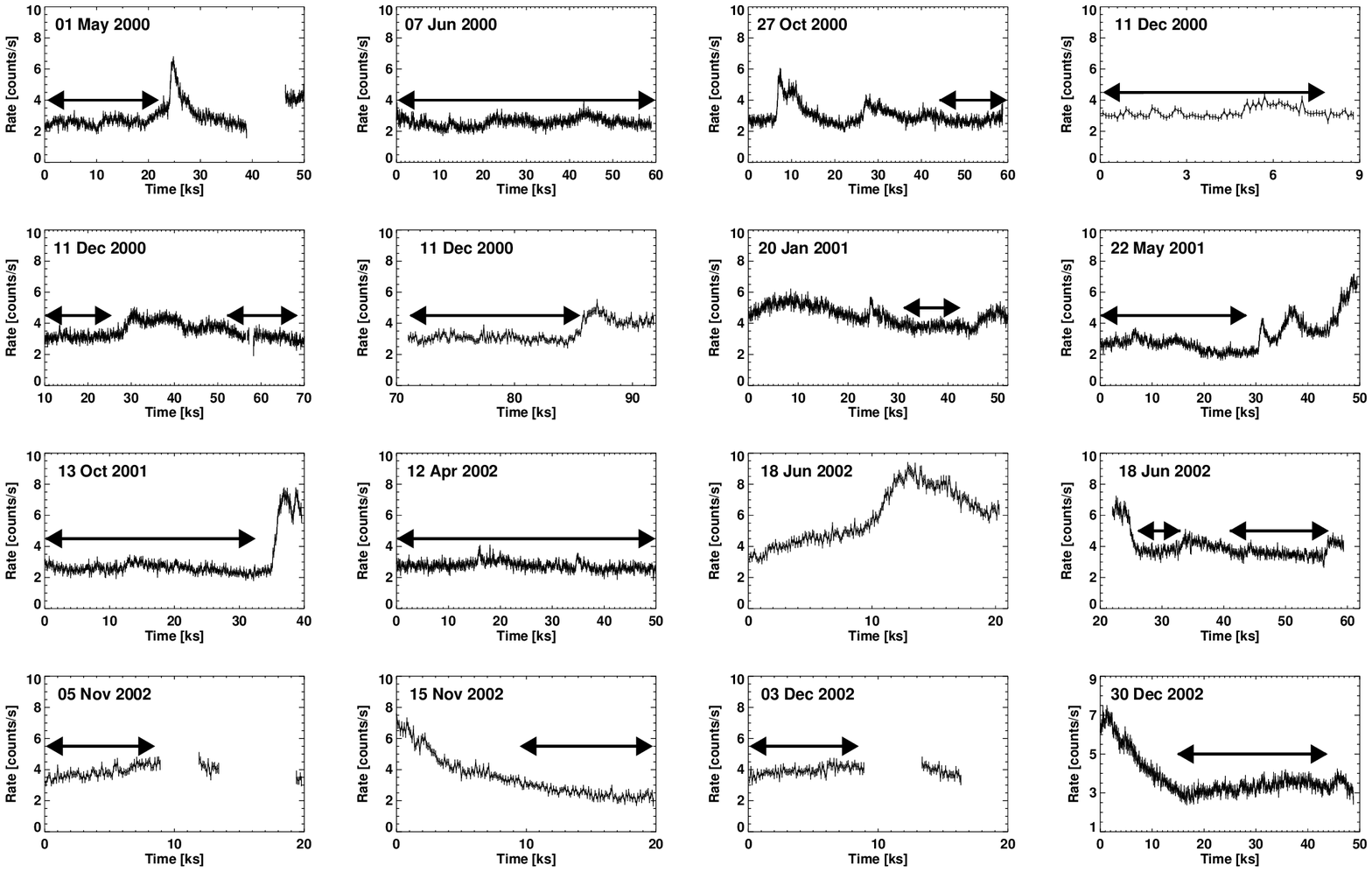}
\includegraphics[width=16.5cm]{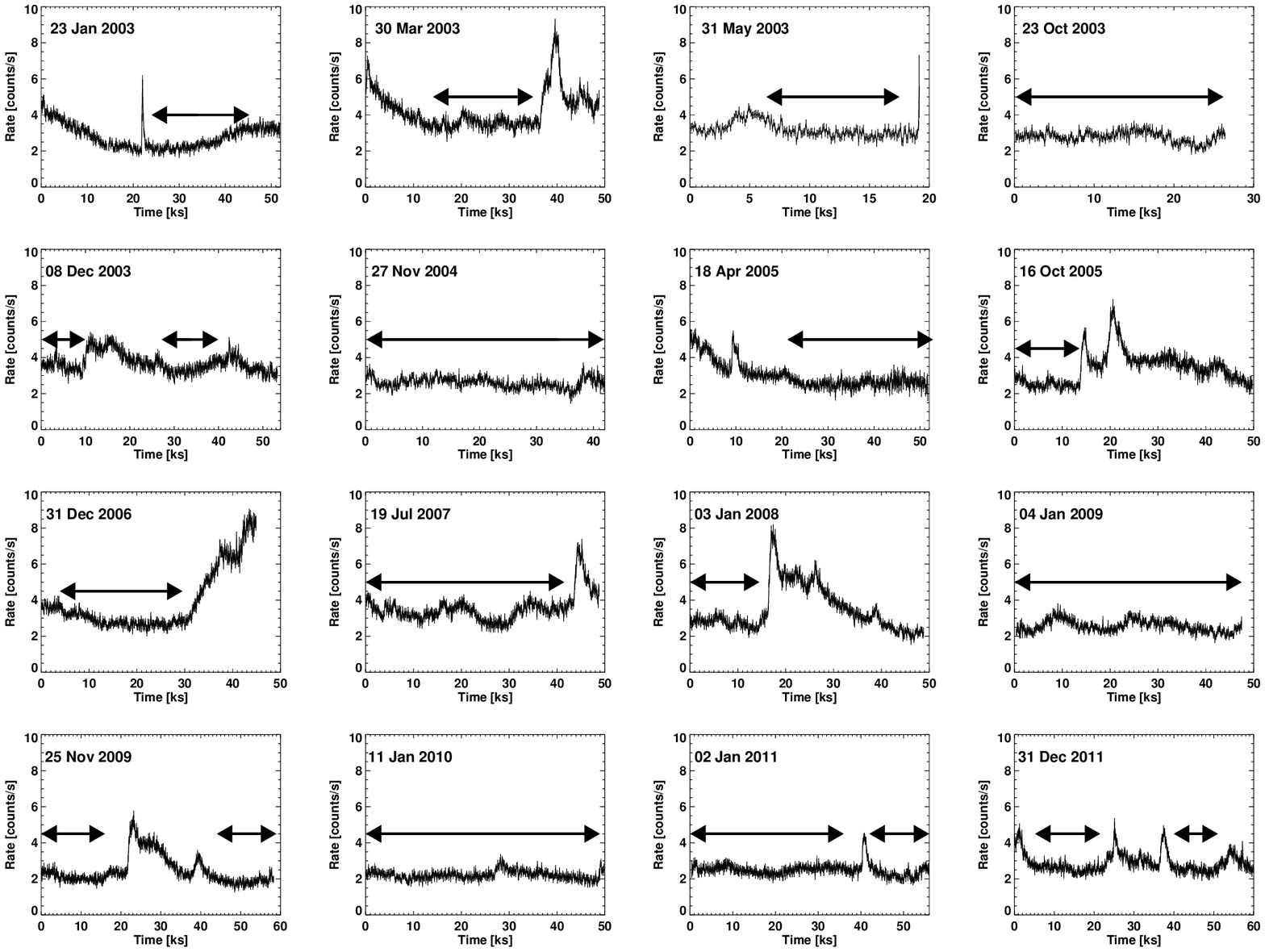}
\caption{\label{flarelc} XMM-RGS light curves of AB~Dor~A plotted in counts per second.  
Quiescent time intervals are marked by arrows; see text for details. A log of the observations is provided in Table~\ref{obs}.}
\end{center}
\end{figure*}

In Fig.~\ref{flarelc}, we provide all RGS light curves\footnote{After revolution 135, 
the CCD 7 in RGS1 suffered a failure;  this failure does affect comparisons between observations using
count rates.} 
used for our analysis, and indicate the times of quiescence and strong flaring. 
Because AB~Dor~A is an active star, flaring is indeed observed in almost all observations. When 
investigating the long-term behaviour of AB~Dor~A, we focused on the quiescent emission. 
Hence we excluded time periods of enhanced activity or strong flaring, particularly 
when the count rate increased from the quiescent level by about 50 or more percent for each observation. 
We thus excluded larger flares on the basis of the respective X-ray light curve and calculated 
the mean or median count rate for the combined RGS (RGS1+RGS2) observations (see Col. 4 in Tab.~\ref{obs}). 

\begin{table}[!ht]
\caption{\label{obs} Observation log of \emph{XMM-Newton} data. Columns 4 and 5 provide 
the mean RGS/median RGS count rate and the data dispersion for the RGS data. }
\fontsize{8pt}{11pt}\selectfont 
\begin{tabular}[htbp]{cccccc}
\hline
\hline
Obs. ID &  Date & Obs. time & Mean/median & $\sigma$ of the\\
&&PN/RGS& RGS count rate & RGS data points\\
&&[ks]&[cts/s]&[cts/s]\\
\hline \\[-3mm]
&2000&&&\\
\hline \\[-3mm]
0123720201& 01/05 & 60.0/49.9&2.64/2.62&0.32\\
0126130201& 07/06 & 41.9/58.9&2.58/2.57&0.29\\
0123720301& 27/10 & 55.7/58.8&2.82/2.80&0.26\\
0133120701& 11/12 & 6.2/8.8&3.07/3.00&0.18\\
0133120101&   "       & 13.4/60.4&3.15/3.13&0.21\\
0133120201&   "        & 4.2/20.8&3.03/3.03&0.21\\
\hline \\[-3mm]
&2001&&&\\
\hline \\[-3mm]
0134520301& 20/01 & 48.6/52.2&3.81/3.80&0.21\\
0134520701& 22/05 & 48.2/49.5&2.58/2.57&0.38\\
0134521301& 13/10 & --- ~/39.6&2.76/2.74&0.26\\
\hline \\[-3mm]
&2002&&&\\
\hline \\[-3mm]
0134521501& 12/04 & 15.9/53.1&3.63/3.66&0.28\\
0155150101& 18/06 & 4.9/20.3&3.74/3.68&0.36\\
0134521601&   "	& 21.3/47.8&3.80/3.77&0.350\\
0134521801& 05/11 & --- ~/19.8&3.86/3.80&0.11\\
0134521701& 15/11 & --- ~/19.8&2.50/2.50&0.32\\
0134522001& 03/12 & --- ~/22.3&3.93/3.91&0.28\\
0134522101& 30/12 & --- ~/48.8&3.31/3.31&0.30\\
\hline \\[-3mm]
&2003&&&\\
\hline \\[-3mm]
0134522201& 23/01 & --- ~/51.8&2.17/2.19&0.17\\
0134522301& 30/03 & --- ~/48.8&3.64/3.59&0.38\\
0134522401& 31/05 & --- ~/28.8&2.95:2.97&0.22\\
0160362701& 23/10 & --- ~/26.5&2.79/2.80&0.23\\
0160362801& 08/12 & --- ~/53.7&3.67/3.63&0.30\\
\hline \\[-3mm]
&2004&&&\\
\hline \\[-3mm]
0160362901& 27/11 & --- ~/56.3&2.60/2.59&0.28\\
\hline \\[-3mm]
&2005&&&\\
\hline \\[-3mm]
0160363001& 18/04 & --- ~/52.1&2.57/2.58&0.38\\
0160363201& 16/10 & --- ~/50.1&2.54/2.51&0.25\\
\hline \\[-3mm]
&2006&&&\\
\hline \\[-3mm]
0412580101& 31/12 & --- ~/44.9&2.96/2.87&0.40\\
\hline \\[-3mm]
&2007&&&\\
\hline \\[-3mm]
0412580201& 19/07 & --- ~/48.8&3.32/3.35&0.42\\
\hline \\[-3mm]
&2008&&&\\
\hline \\[-3mm]
0412580301& 03/01 & --- ~/48.8&2.78/2.77&0.27\\
\hline \\[-3mm]
&2009&&&\\
\hline \\[-3mm]
0412580401& 04/01 & 47.0/48.8&2.56/2.53&0.32\\
0602240201& 25/11 & 57.9/58.3&2.15/2.12&0.23\\
\hline \\[-3mm]
&2010&&&\\
\hline \\[-3mm]
0412580601& 11/01 & 9.9/49.8&2.21/2.20&0.24\\
\hline \\[-3mm]
&2011&&&\\
\hline \\[-3mm]
0412580701& 02/01 & 9.9/62.8&2.51/2.522&0.21\\
0412580801& 31/12 & 9.9/61.8&2.68/2.67&0.22\\

\hline
\end{tabular}
\end{table}

\subsection{Optical data}

We compiled all publicly available photometric V-band data of AB~Dor~A, covering nearly 34 years of 
observations taken between 1978-2012 with a short gap between 1998-1999 and 2000-2001. 
The data taken between 1978-2000 have been presented by \cite{jarvinen_2005}; most 
of the observations were carried out using the standard Johnson \emph{UBVRI} filters. 
Additionally, we used an -- unpublished -- data set collected between 2001-2012 obtained 
in the context of the all-sky automated survey (ASAS)\footnote{ The ASAS data are 
available at http://www.astrouw.edu.pl/asas/} in the V band \citep{pojmanski_1997, pojmanski_2005}, which is publicly available.

ASAS is a CCD photometric sky survey, monitoring 
the southern as well as a part of the northern sky ($\delta < +28^{\circ}$) 
since 2000 up to now. 
The ASAS telescope is located in Chile, Las Campanas Observatory (LCO), at an altitude of 2215~m above sea level and consists of two wide field ($9^{\circ} \times 9^{\circ} $) cameras 
equipped with both V and I filters. For AB~Dor~A, we used observations carried 
out using only V-band data with exposure times of 180s for each frame;
in general, the photometric accuracy of ASAS data for AB~Dor~A is about 0.05 mag.

\section{Long term light curves}\label{sec:lc}

\subsection{Optical light curves}

\begin{figure}
\begin{center}
\includegraphics[width=8cm,height=6cm]{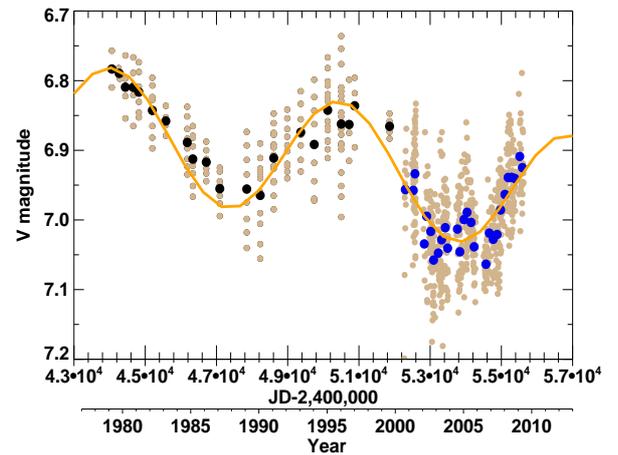}
\caption{\label{optical_lightcurve} AB~Dor~A long-term V-band brightness evolution 
adapted from \cite{jarvinen_2005} and ASAS observations. 
The brown circles denote all individual V-band observations. 
The estimated median magnitudes 
are denoted as black and blue circles for the data from \cite{jarvinen_2005} and the ASAS observations, respectively.
Plotted as a thick line is the sinusoidal fit to the entire dataset 
with a period of $\approx$17 years.}
\end{center}
\end{figure}

In Figure~\ref{optical_lightcurve}, we plot the V-band brightness of AB~Dor~A as a 
function of time. 
We then subdivided the entire $\sim$34 years of V-band observations into smaller time periods 
and estimated a median V magnitude over each of these time bins (depicted as black and blue circles).

\begin{figure}
\begin{center}
\includegraphics[width=8cm]{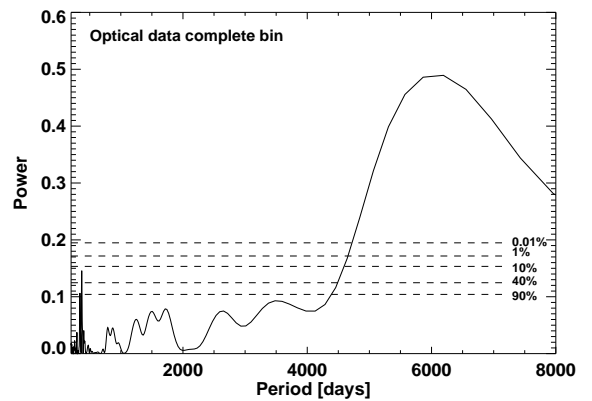}
\caption{\label{period} Periodogram of complete data set of optical V-mag brightness. 
The highest peak indicates activity-cycle period values of 16.96 years.}
\end{center}
\end{figure}

\begin{figure}
\begin{center}
\includegraphics[width=8cm]{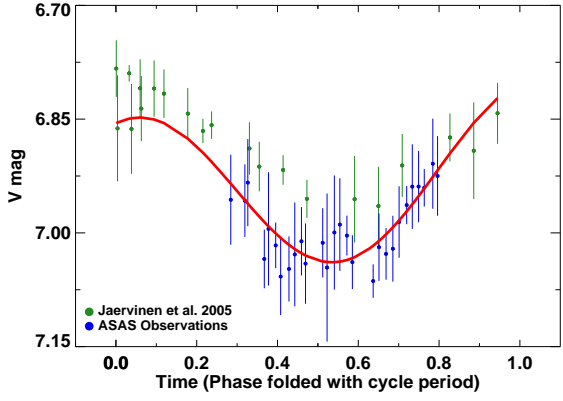}
\caption{\label{opt_folded} Optical V-band brightness data of Fig.~\ref{optical_lightcurve} 
folded with a cycle period of 16.96 years versus the phase interval [0.0,1.0]. Plotted in green and blue are the mean of 
the \cite{jarvinen_2005} and ASAS observations, respectively; the plotted 
error bar depicts the brightness measurement distribution due to rotational modulation. }
\end{center}
\end{figure}

To search for periodic variability 
we performed a periodogram analysis on the entire optical dataset 
using the generalised Lomb-Scargle periodogram 
in the form introduced by \cite{zechmeister_2009},
which is a variant  of the Lomb-Scargle periodogram. 
In Figure~\ref{period}, we show the resulting periodogram  from the 
entire optical data set spanning nearly 34 years of observations.
A clear peak around $\approx$ 6190 days (corresponding to 16.96 years) is 
apparent, which is highly significant, given the derived false-alarm probabilities (FAP), 
also shown in Fig.~\ref{period}.
As a next step we fitted a sine wave with a period of $\sim 16.96$ years 
to the the entire data set presented in  Fig.~\ref{optical_lightcurve}
after correcting for the linear trend in the 
\cite{jarvinen_2005} and the ASAS data set (plotted as orange curve in Fig.~\ref{optical_lightcurve}).
Additionally, in Fig.~\ref{opt_folded} we plot the mean of optical data folded with the 
cycle period of $\approx$ 17 years.

In addition to the main peak in Fig.~\ref{period}, we note another peak 
with a period of $\approx$1 year that has a FAP$\approx 10$\%. 
To determine whether this period is due to the activity cycle we recomputed 
the periodogram after subtracting the best-fit sine wave with
P~=~16.96~years from the observed data.
When comparing both the periodograms it became evident that the high peak with 
a period of $\approx 1$ year persists, hence this peak cannot be the result of considerable spectral leakage
from the cycle frequency.
Therefore, the nature of this peak remains unclear, but we assume that it is caused by the seasonal distribution of the observations.

\subsection{X-ray light curves}

\cite{lockwood_1997} and \cite{radick_1998} studied the relationship between the 
photometric variability and chromospheric activity of Sun-like stars by combining the Mount Wilson HK activity observations 
with  11~years of Str\"omgren b and y photometry taken at Lowell Observatory. 
The latter authors found that on cycle time-scales 
young active stars show an inverse correlation between photometric brightness and 
chromospheric activity, while older stars such as the 
Sun show a direct correlation between brightness and activity. \cite{radick_1998} explained this finding by 
arguing that the stars switch from spot-dominated to facular-dominated 
brightness variations at an age of $\approx$2~Gyr. If the same pattern of variation applies to AB~Dor~A, one would 
expect spot-dominated brightness variation, since AB~Dor~A is a very young star with an age of $\sim50$~Myr \citep{close_2005}. 
Hence, according to Fig.~\ref{optical_lightcurve}, 
one would expect an increase in X-ray activity from minimum to maximum between 1996-2004, 
and a decline in X-ray activity between 1990-1996 and also since 2005. 
In this section we therefore investigate to what extent the available X-ray data support the view of such cyclic coronal activity in AB~Dor~A.

\subsubsection{Overall X-ray behaviour}\label{}

\begin{figure}[h]
\begin{center}
\includegraphics[width=8cm,height=6cm]{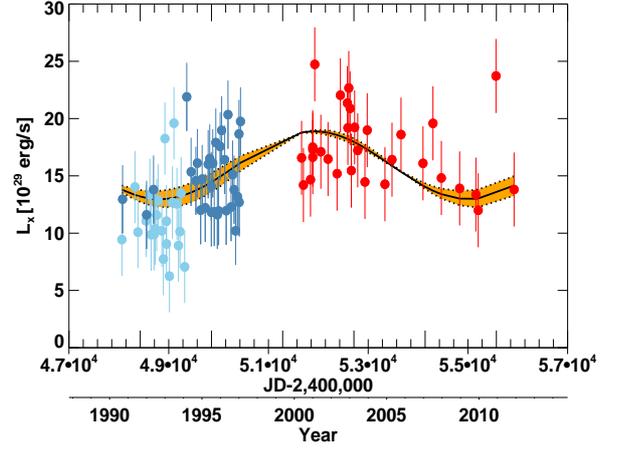}
\caption{\label{lightcurve} Temporal behaviour of the soft X-ray luminosity with 
1$\sigma$ deviation as observed by several X-ray missions between 1990 and 2011. 
\emph{ROSAT} PSPC data are plotted as light-blue filled circles; \emph{ROSAT} HRI data are depicted as navy-blue filled circles. 
The red circles
represent \emph{XMM-Newton} RGS observations. Plotted as a thick black curve is sinusoidal 
fit to the X-ray data with an optical-cycle period of $\approx$17 years.}
\end{center}
\end{figure}

In Figure~\ref{lightcurve}, we show the temporal behaviour of the 
soft X-ray luminosity as observed between 1990 and 2011 by various X-ray satellites.  
Assuming that AB~Dor-A's X-ray spectrum can be described with temperature components 
2 and 5~MK and an equivalent absorption column $N_H$ of $10^{18}$~cm$^{-3}$, 
we computed energy conversion factors (ECF) to convert the observed count rates into fluxes. 
Using XSPEC v12.6.0 and WebPIMMS v4.6, we estimated $ECF_{PSPC}$=$6.42\times10^{-12}$erg/cm$^{2}$/counts, 
$ECF_{HRI}$=$3.03\times10^{-11}$erg/cm$^{2}$/counts, 
$ECF_{RGS1}$=$4.44\times10^{-11}$erg/cm$^{2}$/counts, and $ECF_{RGS2}$=$5.68\times10^{-11}$erg/cm$^{2}$/counts
in the canonical ROSAT energy band 0.1-2.4~keV; the resulting X-ray luminosities (L$_X$) 
were finally calculated using a distance of 14.9$\pm$0.1 pc \citep{guirado_2011}.

We performed a periodogram analysis on the X-ray light curve presented in Fig.~\ref{lightcurve}, 
but we were unable to obtain a significant peak indicating any preferred period. 
Subsequently, we used the period obtained from the optical light curve assuming that the optical
period also applies to the X-ray data. We folded the 
X-ray light curve with this period and obtained a sinusoidal fit to the X-ray data (depicted as a black thick line). 
While this sinusoidal curve 
provides a description of the data, it is far from unique and there is very large scatter
around the fit curve, casting some doubt on signatures of cyclic activity in the X-ray range.

\subsubsection{Correlation between X-ray and optical data}

\begin{figure}[!ht]
\begin{center}
\includegraphics[width=8.0cm]{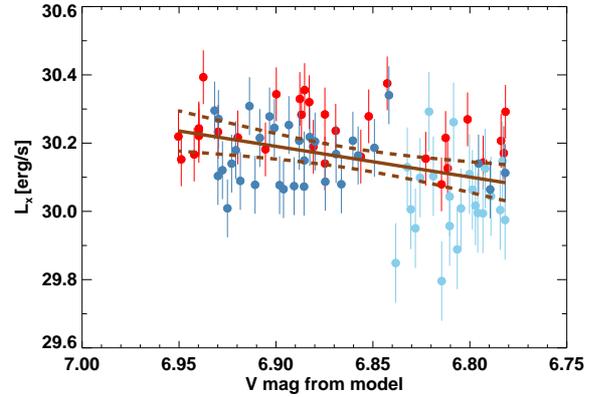}
\caption{\label{hypo} 
Variation of X-ray luminosity 
as a function of the (calculated) V magnitude. The symbols here are the same as in Fig.~\ref{lightcurve}; see text for details.
Plotted as brown lines are the linear regression with 2$\sigma$ confidence band.}
\end{center}
\end{figure}

To examine whether the trends in the X-ray and optical are really correlated, 
we carried out a correlation analysis of the two data sets. To relate an 
optical magnitude to each X-ray observation, we used the value of the fitted optical light curve (see Fig.~\ref{optical_lightcurve}) 
at the time of each X-ray observation. The resulting scatter plot is shown in 
Fig.~\ref{hypo}, where we show logarithmic X-ray luminosity vs. V-band magnitude.
A linear fit between those quantities gives a slope of 0.9$\pm$0.2 for the correlation of log~L$_X$ with V$_{mag}$,
implying that the X-ray luminosity is higher when the photospheric brightness is lower.
Furthermore, we also computed a linear Pearson correlation coefficient ($\rho$) of 0.40 between the 
X-ray luminosity and photospheric brightness with a two-tailed probability value of 0.0001$\%$. 
These findings are clearly consistent with the picture that the star is X-ray bright when the surface brightness is low.

\subsubsection{Individual light curves: ROSAT}

While the overall X-ray light curve might be affected by 
errors in the instrumental cross calibrations, 
trends in individual instruments should be free from such effects and be real.
In the following we therefore concentrated on the data taken with the \emph{ROSAT} PSPC and the 
\emph{ROSAT} HRI, from which a multitude of observations is
available, to determine count-rate trends individually for each instrument. 
We found a negative slope with respect to the time evolution of the count rates for the 
ROSAT PSPC data, while for the 
ROSAT HRI data we found a faint positive slope (see column 2 in Table~\ref{table1}).

We used a simple bootstrap technique to estimate the error of the slope. 
For this purpose we ran a Monte Carlo simulation for the observing times, 
carrying out linear fits to the simulated data sets. The count rates and their individual errors 
were randomly redistributed over the range of available observing times.
A regression analysis on the re-sampled data was performed and 
repeated several times (5$\times$10$^5$ times), thus providing an error at 
$1\sigma$ probability associated with the determined slope.
The results of this analysis are listed in column 3 of Table.~\ref{table1}. 
For the PSPC data we found that in 72$\%$ of the cases slopes as small as the
observed one are obtained by pure chance, whereas for the HRI data this is estimated to be true in 61$\%$ of the cases.

Comparing the optical light curve 
(Fig.~\ref{optical_lightcurve}) and the X-ray light curve (Fig.~\ref{lightcurve}), 
one expects the ROSAT PSPC observations to be at activity minimum during the 17-year cycle, the ROSAT HRI observations 
to be during the constant or rise phase from minimum to maximum activity level and the 
\emph{XMM-Newton} observations cover almost half an activity-cycle period. 
Our regression analysis of the individual PSPC and HRI data is consistent 
with this picture, but the statistical significance is low.

\begin{table}
\begin{center}
\caption{\label{table1} Results from the search for
       a long-term variation in the X-ray data. The errors and the false-alarm probability (FAP) are obtained from 
       bootstrapping the observed distribution of the measurements (BS).}

\begin{tabular}[htbp]{ccccc}
\hline
Data set   & best fit slope  & Error & FAP\\
& [cts/sec/yr]   &  BS  & \% \\
\hline
\emph{ROSAT} & & & \\
PSPC & -0.22  & 0.47 & 72 \\
HRI  & ~0.05  &  0.05 & 61\\
\emph{XMM}
RGS    & -0.04 & 0.02 & 63\\
\hline

\end{tabular}

\end{center}
\end{table}

\subsubsection{Individual light curves: \emph{XMM-Newton}}

\begin{figure}[h]
\begin{center}
\includegraphics[width=9cm]{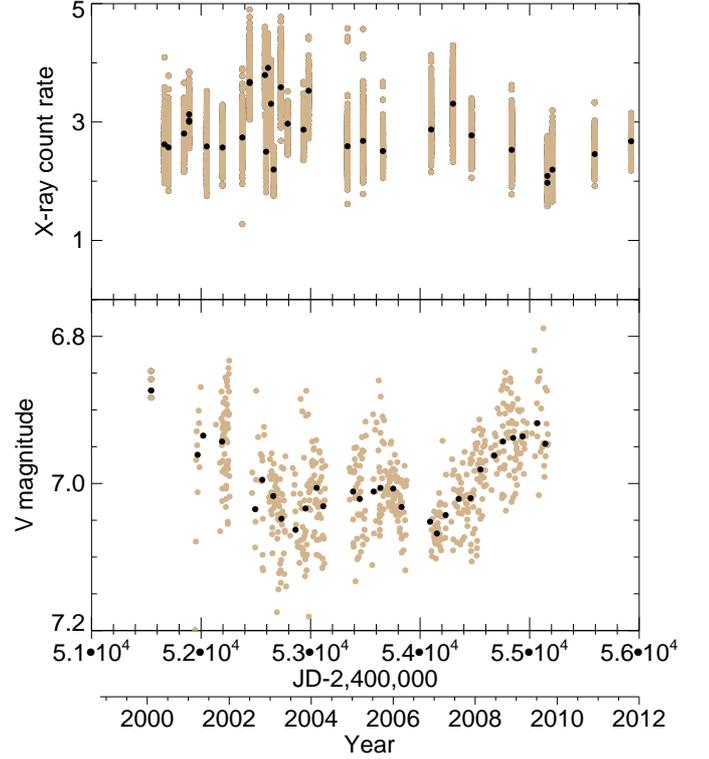}
\caption{\label{cts_lc} Top panel: Temporal behaviour of the XMM-RGS count rate  binned to 100~s 
(as brown circles) after removing the flares from each observation. 
The median 
count rate over the duration of each observational run is depicted as black circles. 
Bottom panel: V-band data for AB~Dor~A (same as Fig.~\ref{optical_lightcurve}).}
\end{center}
\end{figure}

Since AB~Dor~A was used as a calibration source for 
\emph{XMM-Newton}, 
many datasets with much enhanced quality and long temporal coverage of AB~Dor~A have become available, 
which can be used for cycle studies. In contrast to the ROSAT observations, those of XMM-Newton are much longer, which allows us to
identify periods of flaring in the data stream and exclude these periods from analysis. 
In Fig.~\ref{cts_lc}
we plot the evolution of the XMM-RGS count rate taking into consideration only the 
quiescent emission. 
In addition, we re-plot Fig.~\ref{optical_lightcurve} to 
compare the optical and X-ray light curves (the lower panel of Fig.~\ref{cts_lc}). 
If the XMM-RGS data show a variation similar to the optical data, 2000-2006 should represent the activity maximum. 
Because visual inspection suggests 
an anti-correlation between the optical and the X-ray data, we  
carried out some statistical tests to determine whether these trends are significant or not. 

A (parametric) regression analysis on these data similar to the ROSAT data was performed, and the results are presented 
in Table~\ref{table1} as well. We obtained a negative slope for the observed XMM-RGS count rates as a function of time, 
and again, similar to the ROSAT data, we performed a simple bootstrap technique to estimate the error 
of the slope and FAP. A slope as small as the observed one for the XMM-RGS data that occurs by pure chance is estimated to be 63$\%$.  
The negative slope fits with the overall picture of an expected
decline in the activity, but the significance of this slope is again very low.

We then decided to apply non-parametric correlation tests such as the Spearman $\rho$ and Kendall 
$\tau$ test \citep{press_1992} on the mean and the median XMM-RGS data presented in Table.~\ref{obs}.
For this purpose we divided the XMM-RGS data into two subsets, one covering the years 2000-2005, that is, towards the
anticipated maximum, and the second set between 2005-2011 with declining activity.   
The Spearman rank correlation coefficient $\rho$ is defined as 

\begin{equation}
\rho = \frac{\sum (R_i - \bar{R})(S_i - \bar{S})}{\sqrt {\sum (R_i - \bar{R})} \sqrt {\sum  (S_i - \bar{S})}},
\end{equation}

\noindent where $R_i$ and $S_i$ are the ranks of time and the minimum count-rate values 
respectively. 
The significance of a non-zero value of $\rho$ is computed from a parameter $t$ defined as

\begin{equation}\label{tval}
t=\rho \sqrt{\frac{N-2}{1-\rho^2}},
\end{equation}
\noindent where $\rho$ is the Spearman rank and N is the sample size.  
Note that the significance $t$ is distributed approximately as a Student 
distribution with N-2 degrees of freedom \citep{press_1992}; 
a low value of significance ($p$ in Tab.~\ref{table2}) indicates a significant correlation. 
 
An alternative non-parametric test is the Kendall $\tau$-test, which
uses the relative ordering of the rank instead of the numerical difference of the ranks. 
Consider two samples (with n items each) of physical quantities, in our case the observing times $t_i$ ($i=1,N$) 
and the minimum count rate $r_i$  ($i=1,N$); we assumed the times to be sorted so that
$t_i < t_{i+1}$. The total number of possible pairs of time and count rates is n(n-1)/2. We considered 
a pair of values of time $t$ and count rate $r$.  
If the relative ordering of the ranks of two observing times is the same as the relative ordering of the two rates, 
the pair is called concordant, otherwise the pair is called discordant.  Ignoring the problem of how to treat 
timed observations, the basic idea is to compare the number of concordant and discordant pairs, since that number should be statistically
equal in the absence of correlations.  
Specifically, the Kendall $\tau$ is given by 
\begin{equation}
\tau= \frac{n_c - n_d}{n(n-1)/2},
\end{equation}

\noindent 
where n$_c$ is the number of concordant and n$_d$ the number of discordant pairs, 
normalised by the total number of pairs.   Clearly, for a perfect correlation $\tau = 1$.
On the null hypothesis of independence of time and count rate, that is, no correlation, 
$\tau$ is expected to be normally distributed with zero expectation and a variance of 

\begin{equation}
Var(\tau)= \frac{4N+10}{9N(N-1)}.
\end{equation}

The results of our correlation analysis are provided in Tab.~\ref{table2}, where we also quote 
the two-sided probability value (p-value) for a given t-value (Eqn. \ref{tval}).
Clearly, the two tests on the median and mean RGS data yield no significant correlations
for the 2000 - 2005 data, while they suggest significant correlations for both the
median and mean RGS data between 2000 - 2011.  If only the 2005 - 2011 data are considered, 
the $\tau$-test suggests a significant correlation, while the correlation is marginal at best
using the Spearman rank $\rho$.  In {\bf all} cases, however, there is an anti-correlation, that is,
the X-ray rate is decreasing according to expectation.
These results suggests that there is an influence of the activity cycle on the 
X-ray emission, but the observed X-ray variation is quite different 
from the X-ray variation measured for the Sun. It is of course difficult to assess this 
influence quantitatively, but
inspecting the values provided in Tab.~\ref{obs}, we find a maximum
count-rate of 3.93/3.91~cts/s from XMM-RGS data and a minimum count-rate of 2.15/2.12~cts/s,
from which we calculate a variation amplitude of at most $\sim$1.8 in the X-ray emission between
lowest and highest activity in AB~Dor~A, which is consistent with 
previous findings reported by \cite{sanz_2003}.

\begin{table}
\begin{center}
\caption{\label{table2} Results from the correlation test performed on the mean and median XMM-RGS data. 
Column 3 shows the significance of a non-zero value of the Spearman rank, and 
Column 5 shows whether the observed value of $\tau$ is significantly different from zero.}

\begin{tabular}[htbp]{ccccc}
\hline
Data set   & Spearman $\rho$ test  & Kendall $\tau$ test \\
           &~~~~ $\rho$ ~~~~~~ ~~t~~~ p-value&  $\tau$ ~~~~~ $\sigma^2$\\
\hline
Mean RGS data&&\\
2000-2005 &  -0.01 ~~ -0.07 ~~0.94 &  ~0.02 ~~~0.02  \\
2005-2011 &  -0.45 ~~ -1.33 ~~0.22 &  -0.28 ~~~0.07 \\
2000-2011 &  -0.36 ~~ -2.15 ~~0.04 &  -0.22 ~~~0.01 \\

\hline
Median RGS data&&\\
2000-2005 &  -0.05 ~~ -0.22 ~~ 0.82&  ~0.01 ~~~0.02  \\
2005-2011 &  -0.33 ~~ -0.94 ~~ 0.37&  -0.22 ~~~0.07 \\
2000-2011 &  -0.35 ~~ -2.05 ~~ 0.04 &  -0.20 ~~~0.01 \\
\hline

\end{tabular}

\end{center}
\end{table}


\subsection{Summary: Is there an X-ray activity cycle on AB~Dor~A~? } 

While a clear cyclic behaviour with a cycle length of $\sim$17 years 
is observed for AB~Dor~A in its optical brightness variations,
a similar variation in the available X-ray data is not immediately 
apparent. However, Fig.~\ref{hypo} suggests an anti-correlation between optical and X-ray brightness
in support of the view of a variation in X-ray flux with the optical cycle.
The extensive and contiguous observations carried out with the XMM-Newton RGS allow a much more
refined assessment of the temporal variability of AB~Dor~A than all previously
available X-ray data. 
An inspection of Fig.~\ref{flarelc} demonstrates that
AB~Dor~A is variable at all times and does indeed produce frequent and significant flaring.
However, the  {\it XMM-Newton} RGS data also demonstrate that AB~Dor~A returns to a basal state at around approximately 3 RGS cts/sec.  
This basal state can be observed only in reasonably long and contiguous observations, and even then it may not be attained. 
At any rate, in short and non-contiguous data, as available from satellites in low Earth orbit such as
ROSAT, it is difficult to identify such basal state periods;
still, taking the ROSAT data at face value, the data support a
variation of X-ray flux with optical cycle in the anticipated way, although the obtained
correlations are not statistically significant.

Since the available \emph{XMM-Newton} observations now cover the period between 
optical activity maximum and
minimum, we carried out several statistical tests to study a possible activity cycle 
associated with the XMM-RGS data. The results indeed indicate an increase and 
decline in activity with an activity maximum around 2002-2003 (cf. Fig.\ref{cts_lc}). 
This change in activity manifests itself in a change in the flux of the
basal state level, but the change in amplitude is at most a factor of two, and possibly even lower.
As a consequence, the relative change is far less than the relative change observed in the Sun and
other late-type stars and therefore the variability of the star's basal state is weaker than 
the typical X-ray variability (outside flares) in less active cool stars. The low-amplitude 
variability observed in AB~Dor~A may be attributed to the fact that we are dealing with an ultra-fast rotator that has a saturated 
corona.

\section{Comparison of AB~Dor~A's activity cycle with that of other stars}\label{sec:compare}

\begin{figure}[h]
\begin{center}
\includegraphics[width=8.0cm]{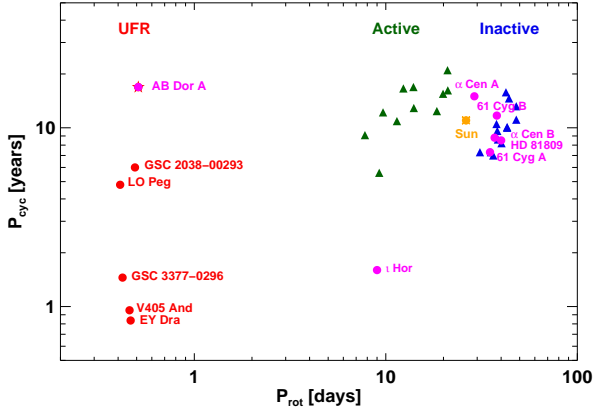}\vspace{-0.4mm}
\caption{\label{cycrot} Rotation period plotted as a function of the activity-cycle period. 
Depicted as blue and green triangles are stars 
belonging to inactive and active sequence \citep{bohm_2007}. 
Represented as a magenta circles are stars with X-ray cycles (cf. see introduction). 
The place of AB~Dor~A is represented 
as magenta star. Depicted as red circles are the activity cycles reported in other ultra-fast rotators reported by 
\cite{bernhard_2006}, \cite{tas_2011}, and \cite{vida_2013}.}
\end{center}
\end{figure}

In the following section we view our findings on the activity cycle on AB~Dor~A in the context of  
stellar activity cycles as seen in X-rays and other activity indicators.
In Fig.~\ref{cycrot}, we plot the cycle period (in years) vs. the stellar rotation period
(in days) for the sample stars.  We use the stellar sample discussed in detail by 
\cite{brandenburg98} and
\cite{bohm_2007} (blue and green triangles), the stars with confirmed X-ray
cycles discussed in the introduction section (magenta circles) 
and individual fast rotators with activity cycles as discussed by \cite{bernhard_2006}, 
\cite{tas_2011}, and \cite{vida_2013} (red circles); the data point for AB~Dor~A is also shown, 
using its 0.52-day rotation period and 17-year activity-cycle period.

\cite{brandenburg98} showed that active and inactive stars follow different branches
in a $P_{cyc}$- $P_{rot}$-diagram. 
\cite{bohm_2007} suggested that the time taken for the toroidal magnetic field to reach the 
stellar surface is determined by the length of the activity cycle associated with 
the star. Hence studying the relation between the rotation 
period and the length of the activity cycle may shed light on the relevant dynamo mechanisms.

Most of the X-ray cycle stars fit the active or inactive sequence
proposed by \cite{brandenburg98} and \cite{bohm_2007} reasonably well. 
Note that the short-period systems follow yet different branches in the $P_{cyc}$- $P_{rot}$-diagram. 
We hypothesise that AB~Dor~A and other ultra-fast rotators have somewhat different dynamo processes that are 
not readily comparable to more slowly rotating stars. Clearly, substantial work
needs to be done to demonstrate the reality of {\it activity cycles} in ultra-fast rotators as a class.

\section{Rotational modulation}\label{sec:rotmod}

\begin{figure}[h]
\begin{center}
\includegraphics[width=8.0cm,height=5.5cm,clip]{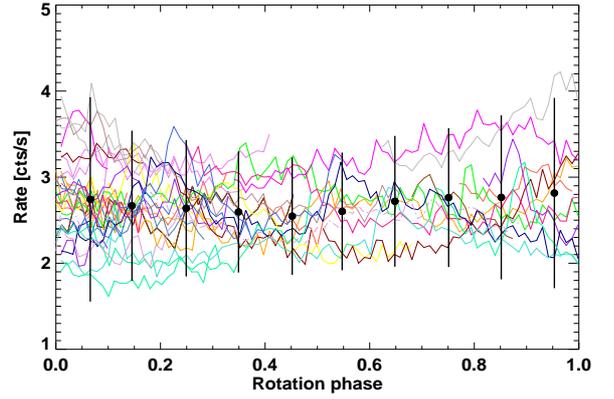}\vspace{-0.4mm}
\caption{\label{rgs_lc} X-ray light curve after removing the flares from a subset of
 XMM-RGS observation on AB~Dor~A folded with rotational period and plotted
vs. the phase interval. Each colour represents different observations/ different rotation. 
Represented as black filled circles are mean count rate at a certain rotation phase with 2$\sigma$ deviation.}
\end{center}
\end{figure}

Out of the wealth of data of available {\it XMM-Newton} data on AB~Dor~A, we chose the
subsets of data that cover more than one stellar rotation and are therefore well suited for a 
short-term variability study. 
In Fig.~\ref{rgs_lc}, we depict the X-ray light curve after flares were removed by eye 
for all XMM-RGS data sets with more than one rotational period and 
plot this vs the rotational phase interval. 
We note substantial fluctuations as a result of an active corona. 
The seemingly irregular variability seen in individual 
light curves can be attributed to the low energy and short time-scale flares, with no
obvious sign of rotational modulation. We point out that the data shown in 
Fig.~\ref{rgs_lc} extend over ten years, yet the dispersion of the data is very low, 
re-emphasizing the existence of a possible basal coronal state in AB~Dor~A.

\section{Summary}\label{sec:sum}

The available X-ray observations of AB~Dor~A, compiled for a period of more than 
two decades, show X-ray variability on a variety of time scales. Since 
AB~Dor~A is a very active star, it exhibits substantial short-term variability and 
in particular frequent flaring activity. 
This flaring activity may last for several hours and can be observed in {\it XMM-Newton} data 
where the exposure time ranges over several tens of kilo-seconds (cf. Fig.~\ref{flarelc}). But in short and 
non-contiguous snapshot observations typically available from 
low Earth-orbit satellites such as the ROSAT observations, which typically have individual 
exposure times of a few kilo-seconds at best, this variability is difficult to distinguish from 
variability on longer time-scales.
In the optical, long-term variability with a period of 
16-17 years that is highly reminiscent
of the solar sunspot cycle could be established; with 20\% the amplitude of the
optical cyclic variations is quite substantial.

Because of its substantial variability on short time-scales, a correlation between X-ray and 
photospheric activity is difficult to establish with snapshot X-ray data. 
However, in sufficiently long X-ray observations of AB~Dor~A, we were able to estimate a basal state of 
$\approx$ 3 {\it XMM-Newton} RGS cts/sec.  Furthermore,  
we presented evidence that this basal state flux may vary with the optical cycle of AB~Dor~A in analogy to the solar activity 
cycle; note that the Sun appears faintest when large Sun spot group(s) are
on its visible hemisphere.
However, for AB~Dor~A we estimated a factor of only 
$\sim$1.8 variation in the X-ray emission in 0.3-2.5~keV range during its cycle, and
therefore this very active star is in some sense much less variable than other
solar-like stars \citep{robrade_2012}, 
albeit the basic picture that the X-ray flux is highest, when the photospheric brightness is at lowest, 
also seems to apply to AB~Dor~A.

Clearly, the time interval for which X-ray and optical data are available is very short compared to
the assumed cycle period of almost 17 years. Only two cycles have been covered so 
far, and a true periodicity as observed for the Sun is far from established. 
Furthermore, photometric monitoring of AB~Dor~A and related objects in the next decades 
is certainly in order to study whether the observed variations are truly cyclic with a well-defined period.  With more or less
robotic facilities such monitoring can today be carried out at relatively low cost.
Similarly, extended X-ray observations of AB~Dor~A and again similar objects would help in establishing
the existence of such basal states in other stars, as well, and also their relationship to stellar
parameters and possible cycle variations.

\begin{acknowledgements}
S.~L. acknowledges funding by the DFG in the framework
of RTG 1351 ''Extrasolar planets and their host stars''.  
\end{acknowledgements}

\bibliographystyle{aa}

\bibliography{paper}

\end{document}